\documentclass[aps,pre,showpacs,preprint]{revtex4}
\usepackage{bm}
\usepackage{graphicx}
\bibstyle{approve.bib}

\begin{document}
\title{On the equilibria of the MAPK cascade: cooperativity, modularity and bistability}
\author{C. Russo$^{1,2}$, C. Giuraniuc$^1$, R. Blossey$^1$ and J.-F. Bodart$^{1,2}$} 
\affiliation{$^1$ Biological Nanosystems, Interdisciplinary Research Institute 
USR3078, 50 Avenue Halley, F-59658 Villeneuve d'Ascq, 
France\\
$^2$ Laboratory of Division Signal Regulation, EA 4020, Building
SN-3, University of Sciences and Technology Lille, 59655 Villeneuve
d'Ascq, France}

\date{\today}

\begin{abstract}
In this paper we present a discussion of a phenomenological model of the MAPK cascade 
which was originally proposed by Angeli {\it et al.} (PNAS {\bf 101}, 1822  (2004)). The model 
and its solution are extended in several respects: a) an analytical solution is given for the cascade equilibria, exploiting a parameter-based symmetry of the rate equations; b) we discuss the cooperativity (Hill coefficients) of the cascade and show that a 
feedforward loop within the cascade increases its cooperativity. The relevance of this result
for the notion of modularity is discussed;  c) the feedback model for cascade bistability by Angeli 
{\it et al.}  is reconsidered. We argue that care must be taken in modeling the interactions and
a biologically realistic phenomenological model cannot be too reductionist.
The inclusion of a time-dependent degradation rate is needed to account for a switching of the 
cascade. 
\end{abstract}


\maketitle

\section{Introduction}

The MAPK ({\bf M}itogen-{\bf A}ctivated {\bf P}rotein {\bf K}inase) cascade
is a paradigmatic signaling cascade which plays a crucial role in many
aspects of cellular events \cite{gomperts02}. The biological function of
the cascade is to ensure that an extracellular signal arriving at the cell 
membrane can be transported reliably to targets in the cytoplasm 
and in the nucleus. The functioning of this signaling mechanism is therefore of
obvious interest to biologists, since interventions in the signaling chain permit 
controlling physiological behaviours. From the point of view of a physicist,
it is interesting to understand how different control mechanisms (nonlinearities 
due to cooperative molecular interactions, feedback loops etc.)  shape 
the behaviour of the signaling chain \cite{ferrell96,ferrell01,tyson03}.

The MAPK cascade consists of several biochemical reaction levels in which 
an activated kinase at each level phosphorylates a kinase at a downstream level:
the signaling chain can thus be considered as a cascade of phosphorylations. 
The MAPKK kinase - the activator at cascade entry -  is the protein Mos in 
{\it Xenopus} oocytes, which underlies  our work (see Figure 1). These amphibian 
gametes are considered as model cells, easy to process experimentally 
due to their size ($\sim$mm), rich in the concentration of the signaling molecules 
(even up to 50 ng of protein per oocyte \cite{sagata88}), and hence signaling events 
can be reliably modeled by ODE-models for the reaction kinetics.

Understanding the control mechanisms involved in the MAPK signaling cascade has 
been a major research topic in recent years in the emerging field of systems biology, 
giving evidence that several aspects of the cascades can be both studied {\it in vivo} 
and modeled mathematically. The key work in this respect was the paper by Huang and 
Ferrell \cite{huang96} which developed a detailed kinetic model for the so-called 
ultrasensitivity of the cascade, based on experimental data obtained
for the cascade in {\it Xenopus}. The notion of `ultrasensitivity' refers to the steepness of 
the sigmoidal  stimulus/response-curve which is the hallmark of a signaling cascade: 
a very nonlinear response ensures an effective response to a 
signal received by the cell.  

In the context of the transformation of a gradual stimulus into a 
nonlinear response curve, the question of the control of the cascade
has attracted the interest of experimenters and theorists alike. In fact, 
depending on how the cascade is `implemented' within other network elements,
it can function as a switch, an amplifier or feedback-controller
\cite{bluethgen01,markevich04,gunawardena05}. This has led to the
view that the cascade can be considered as a `module' inserted in a network
of interactions. In particular, the role of positive and
negative feedback loops in conjunction with the cascade is
a topic of continuing interest \cite{kholodenko00,xiong03}.  
More recently, also the role of stochastic effects in the cascade have been
considered \cite{wang06}, as well as the spatial
progression of the signaling cascade from the plasma membrane to the
nucleus \cite{markevich06}.  

In order to be able to develop increasingly quantitative comparisons
between experiment and models, it is useful to understand the properties
of the models as well as possible; a recent prime example in this context is
the work by Ventura et al. \cite{ventura08}. Due to the complexity of the systems
addressed, it is in fact rare that analytical calculations can be made.
With this work we start a series of investigations on phenomenological
models for the MAPK cascade.  First, we present an exact treatment of the 
MAPK cascade equilibria based on a phenomenological model introduced in ref.
\cite{angeli04}.
 
Here we show a) that for the case of a Michaelis-Menten kinetics, the equilibria
of the cascade can be computed analytically by exploiting a parameter-based 
symmetry of the equations. (For a simple linear
kinetics, this is a text book matter, see \cite{alon06}.) 
We then further address the two following issues:
b) how can the cooperativity of the cascade be changed? We show that by
an indirect feedforward loop, the input signal - response curve of the cascade
can become more pronounced, i.e. acquire a higher effective Hill coefficient;
this results  has repercussions on the notion of modularity of the cascade.
Finally, c), we reconsider results by Angeli {\it et al.} on the feedback-induced bistability
of the cascade and introduce a simple biological mechanism by which the 
cascade can switch off. 

 \section{A Michaelis-Menten model of the MAPK cascade}

Figure 2 presents the basic scheme of the interactions in the cascade 
as we discuss it here. The cascade is initiated by Mos, of which the 
concentration $x$ is taken first as given (i.e., we neglect the kinetics of Mos for 
the moment). Mos activates the phosphorylations of MEK, $y_1$, whose 
once- and doubly-phosphorylated forms are denoted by $y_2$ and $y_3$,
respectively. The second level of the cascade is a repetition of
the first level in which now, however, the role of Mos is played
by doubly-phosphorylated MEK, $y_3$. The second layer of the cascade
is the sequence of phosphorylations of MAPK, denoted by $z_1$,
leading to $z_2$ and $z_3$ in complete analogy to the MEK-level
of the cascade (Figure 2 A). Indicated in Figure 2 B) is the case in which 
Mos can also act within the cascade by promoting phosphorylation of 
downstream targets in an indirect way \cite{verlhac00}. Finally, 
Figure 2 C) shows that the cascade output, 
$z_3$, can act back on the input level $x$ by way of a feedback loop and 
hence affect the concentration of Mos - we will turn to this case in the 
discussion of bistability. 

Following \cite{angeli04} we write the kinetic equations for the system 
as follows
\begin{eqnarray} \label{angeli}
\dot{y}_1 & = & \frac{V_6y_2}{K_6 + y_2} - \frac{V_3xy_1}{K_3 + y_1} \\
\nonumber \\
\dot{y}_2 & = & - (\dot{y}_1 + \dot{y}_3)\\
\nonumber \\
\dot{y}_3 & = & \frac{V_4xy_2}{K_4 + y_2} - \frac{V_5y_3}{K_5 + y_3} \\
\nonumber \\
\dot{z}_1 & = & \frac{V_{10}z_2}{K_{10} + z_2} - \frac{V_7y_3z_1}{K_7 + z_1} \\
\nonumber \\
\dot{z}_2 & = & -(\dot{z}_1 + \dot{z}_3) \\
\nonumber \\
\dot{z}_3 & = & \frac{V_8y_3z_2}{K_8 + z_2} - \frac{V_9z_3}{K_9 + z_3}
\end{eqnarray}
where the $V_i$ and $K_i$ are reaction speeds and equilibrium constants,
respectively, and where the numbering of the reactions follows the scheme by
\cite{kholodenko00} which is also used in ref. \cite{angeli04}. This scheme
simply numbers the reactions sequentially layer by layer, first all phosphorylations,
and then all dephosphorylations. 

Eqs.(2) and (5) are a consequence of the conservation of the total number of the 
proteins MEK and MAPK, i.e.,
\begin{equation}
\sum_{i=1,2,3} y_i = y_T\,,\,\,\,\,\,\sum_{i=1,2,3} z_i = z_T\,
\end{equation}
Thus, $ \dot{y}_2  = - (\dot{y}_1 + \dot{y}_3) $, and likewise for $\dot{z}_2$.
The system of kinetic equations hence reduces to four independent equations only.

We now show that the equilibria of the MAPK-cascade can be determined analytically 
in an exact way; given that the kinetics is nonlinear, this is a non-trivial result. 
To demonstrate it we begin with the equation for $\dot{y}_1$ (i.e., non-phosphorylated MEK),
eq.(1).
Dividing this equation by $V_6$ and redefining the variables via
\begin{equation}
x' \equiv \frac{V_3}{V_6}x\,,\,\,\,\, y_i \equiv y_i'y_T\,,\,\,\, i=1,2,3
\end{equation}
we find
\begin{equation}
\frac{y_T}{V_6}\dot{y}_1' =  
\frac{y_2'y_T}{K_6 + y_2'y_T} - x'\frac{y_1'y_T}{K_3 + y_1'y_T}\, .
\end{equation}
The kinetic parameters and reaction speeds are as estimated
in the Supplementary Material of ref. \cite{angeli04}. Although not exact, $K_3 = K_6 = y_T $, 
is a reasonable assumption which we follow throughout this paper. Making use of 
this symmetry, the equation then simplifies to
\begin{equation}
\frac{y_T}{V_6}\dot{y}_1' =
\frac{y_2'}{1 + y_2'} - x'\frac{y_1'}{1 + y_1'}\, .
\end{equation}
What remains is to redefine time according to $t' = (V_6/y_T)t $.
Dropping the primes we obtain
\begin{equation}
\dot{y}_1 = \frac{y_2}{1 + y_2} - x\frac{y_1}{1 + y_1}\, .
\end{equation}
Due to the further parameter symmetries deduced from experiment by Angeli {\it et al.}, 
$(V_3 = V_4, V_5 = V_6,K_3 = K_4 = K_5 = K_6 = y_T)$, the same procedure also works
for the equations for $\dot{y}_2$ and $\dot{y}_3 $.

In fact, the same strategy can also be applied to the equations for $z_i$, 
$i=1,2,3$. The only differences now are that we have already redefined the
variable $y_3$ and time $t$; this means that the equations for $z_i$
do have to contain two additional parameters which reflect the different
timescales of the dynamics for $y_i$ and $z_i$, and the concentration 
scales of the $ y_i $ and $z_i$. These two new parameters are given by
\begin{equation}
v \equiv \frac{V_7}{V_{10}}y_T\,,\,\,\,\, 
\tau \equiv \frac{V_6}{V_{10}}\frac{z_T}{y_T}\, .
\end{equation} 
For example, the equation for $z_1$ reads
\begin{equation}
\tau\dot{z}_1 = \frac{z_2}{1 + z_2} - v y_3 \frac{z_1}{1 + z_1}\, ,
\end{equation}
and as before, due to the parameter symmetries $(K_7 = K_8 = K_9 =
K_{10} = z_T, V_7 = V_8, V_9 = V_{10})$, the same procedure can
be applied to the equations for $z_2 $ and $z_3$.

We have been able to rewrite the system of four equations (plus two
which are the trivial consequence of the conservation laws) with two 
parameters only, whereby only one of them affects the time course
of the cascade. This is clearly possible in an exact way only for the 
parameter symmetries uncovered by Angeli {\it et al}. We believe,
however, that this does not constitute a major restriction for our intention
to perform analytic calculations. In case the parameters do deviate from
the exactly symmetrical values, our calculation can be used to develop
an expansion in the perturbed parameters.

In a subsequent step we can now profit from the fact that the equations
have acquired a homogeneous form in the variables
\begin{equation}
w_i \equiv \frac{y_i}{1 + y_i}\,,\,\,\,\,\, q_i \equiv \frac{z_i}{1 + z_i}\, .
\end{equation}
Noting further that the time derivative, e.g. of $w_i $, is given by
\begin{equation}
\dot{w}_i = \frac{\dot{y}_i}{(1 + y_i)^2}\, ,
\end{equation}
and similar for the $\dot{q}_i$, we find that the model can be cast
into the simple form
\begin{eqnarray}
\dot{w}_1 & = & (1-w_1)^2 (w_2 - xw_1)\\
\nonumber \\
\dot{w}_2 & = & (1-w_2)^2 (xw_1 + w_3 - xw_2 - w_2)\\
\nonumber \\
\dot{w}_3 & = & (1-w_3)^2 (xw_2 - w_3)\\
\nonumber \\
\tau \dot{q}_1 & = & (1-q_1)^2 (q_2 - s q_1)\\
\nonumber \\
\tau \dot{q}_2 & = & (1-q_2)^2 (sq_1 + q_3 - q_2(1+s))\\
\nonumber \\
\tau \dot{q}_3 & = & (1-q_3)^2 (s q_2 - q_3)
\end{eqnarray}
where
\begin{equation}
s \equiv v\frac{w_3}{1 - w_3}\, .
\end{equation}

From these equations, the fixed-point conditions $\dot{w}_i = \dot{q}_i = 0 $ 
can now easily be read off.
Note that since $w_i \leq 1/2 $, zeroes can only appear in the right-most
bracket of each equation. For the $w_i$ one has
\begin{equation}
w_3 = x w_2 = x^2 w_1 
\end{equation}
which expresses the character of the phosphorylation mechanism in a 
very clear way (and in fact, analogously to the linear kinetics
\cite{alon06}). Similarly, for the $q_i$
\begin{equation}
q_3 = s q_2 = s^2 q_1\, .
\end{equation}

In order to determine the fixed-point values of these quantities
explicitly, we have to invoke the constraints on $y_i$ and $z_i$,  
which have to be expressed in terms of the new variables. The condition
\begin{equation}
y_1 + y_2 + y_3 = 1
\end{equation}
becomes in terms of $w_i$ 
\begin{equation}
\frac{w_1}{1 - w_1} + \frac{w_2}{1 - w_2} + \frac{w_3}{1 - w_3} = 1\, .
\end{equation}
Putting in the fixed-point conditions for $w_2$ and $w_3$, this 
equation turns into a cubic equation for $w_1$,
\begin{equation}  \label{conservation_w1}
4x^3w_1^3 - 3 x (1+x+x^2) w_1^2 + 2 (1 + x +x^2)w_1 - 1 = 0\,.
\end{equation}
Due to the symmetry of the first and the second level of the
cascade, the corresponding equation for $q_1$ is obtained by replacing 
$w_1 $ by $q_1$, and $x$ by $s$.

Clearly, the cubic equations for $w_1$ and $q_1$ can be solved exactly
with textbook formulae. There is a unique real solution which fulfills the 
condition $w_1(x=0) = 1/2$, as follows from a study of
eq.(\ref{conservation_w1}) near $x=0$.
Since for $x \rightarrow 0$, $w_1 \rightarrow 1/2$, we can neglect the cubic 
and quadratic terms and find
\begin{equation} \label{w_1}
w_1 = \frac{1}{2}\frac{1}{1 + x + x^2}\,,
\end{equation}
which is a result analogous to the linear kinetics. Figure 3 (top) compares
eq.(\ref{w_1}) with the numerical solution to eq.(\ref{conservation_w1}).
One sees that also for large values of $x$, the full solution asymptotically
approaches the profile given by (\ref{w_1}). We note that the two other solutions
to eq.({\ref{conservation_w1}) are singular for $x \rightarrow 0 $, as can be
seen by neglecting the constant term -1 and solving the remaining
quadratic equation. For increasing values of $x$, the solutions become complex.}

From the fixed-point conditions $ w_2 = xw_1 $ and $w_3 = x^2 w_1$
one immediately obtains from eq.(\ref{w_1})
\begin{equation}
w_2 = \frac{1}{2}\frac{x}{1 + x + x^2}\,,
\end{equation}
and 
\begin{equation}
w_3 = \frac{1}{2}\frac{x^2}{1 + x + x^2}\,.
\end{equation}
Thus we see that while $w_1$ diminishes as a function of $x$, $w_2$ first
rises linearly, and then drops $\propto 1/x $ for large $x$, due to the
depletion for the doubly-phosphorylated form. The concentration of 
doubly-phosphorylated MEK, by contrast, shows exactly the expected
sigmoidal Hill-type profile with a Hill coefficient of two. As far as the
asymptotics of the Hill coefficients is concerned, there is no difference 
between the linear and the Michaelis-Menten kinetics. The precise details
of the concentration profiles are, however, essential in a
comparison to experiment. We illustrate this point here by comparing our 
approximate (linear) and the exact solution of the cubic equation for the 
case of Michaelis-Menten kinetics.

Figure 3 (top) illustrates the difference between the approximate 
solution (linear) and the exact numerically calculated solution of the cubic 
equation for $w_1(x)$. If one tries to fit  the curve to a pure powerlaw $w_1(x) \sim x^{\, b}$ 
over a range of values $ 1.5 \leq x \leq 3 $, which is roughly an experimentally accessible
window for Mos based on the parameters used and experimental 
data for Mos concentrations \cite{angeli04,sagata88}, 
the best fit is obtained for a value of $ b = -1.557 $. Fitting of the curve over a much 
wider range, one order of magnitude in $x$, reveals that the value of $b$ converges to a value of
two; we find $ b = - 1.942 $. This result of Figure 3 (bottom) shows that the attribution of
the Hill coefficient (the highest nonlinearity in the concentration law)
from experiment is difficult since the lower order terms, which 
decay less rapidly, strongly affect the result. 

Due to the formal identity of the equation for $q_1$ with that of $w_1$,
we immediately find the analogus results for $q_1, q_2, q_3$, this time
with $x$ replaced by $s$. If we transform the final result for $q_3$ back
to $z_3$, we find
\begin{equation} \label{z3}
z_3 = \frac{s^2}{2(1 + s + s^2)}
\end{equation} 
which, due to the quadratic dependence of $s$ from $x$ is actually a
sigmoidal function with a Hill coefficient of four, as is common for the
MAPK cascade; see Figure 4. Again, there is no difference in the asymptotic
value of the Hill coefficient between the linear and the exact cubic case.

\section{Increasing cooperativity, breaking modularity}

We now turn to the second question we want to address: how can one increase the
Hill coefficient, i.e. the cooperativity of the cascade? From the foregoing section we have 
found that the Michaelis-Menten MAPK cascade has a maximal Hill coefficient of four, the highest nonlinearity in the denominator of eq.(\ref{z3}). Evidently, the effectivity of the cascade is based on a sufficiently steep rise in concentration of MAPK at threshold. Is there a way to increase this
steepness, i.e. having a still higher cooperativity? 

In order to answer this question it is useful to investigate the impact of possible 
modifications of the cascade. Two cases may be distinguished: (i) additional 
levels of phosphorylation and (ii) rewirings of the cascade by the introduction of feedback 
and/or feedforward loops.

In the first case one can either have more phosphorylation steps within one given level,
or an increase in the number of cascade levels. Although both will lead to an 
increase in cooperativity - increase the Hill coefficient - but this comes at a cost. 

If more phosphorylation levels are introduced, the unphosphorylated and maximally 
phosphorylated states become disfavoured with respect to intermediate states, since the
overall population of these states has thus to go down for a given constant number of proteins. 
More levels thus means favoring the intermediate states, while it is the entry and exit
states that are really important. The function of the intermediate level is just to separate
entry and exit state by separating them from each other - but the separation should not
be too `large', i.e. involve too many states.

On the other hand, an increasing the number of cascade levels requires that another 
molecular partner has to be involved. This would have as a consequence a decrease 
in the `robustness' of the cascade. The notion of robustness is used here in the sense
that building a cascade based on many different molecules increases the probability
of cascade failure.
 
By contrast, feedback and feedforward loops are schemes through which the cascade
can, in principle, increase its cooperativity without increasing the type of molecules 
intervening and maintaining both phosphorylation levels and cascade levels. 

We now show that a feedforward loop in the cascade can indeed lead to an increase 
of cooperativity. Such a loop may be introduced in the cascade when one considers Mos 
to act through two opposite pathways within the cascade. It has been observed that 
the MAPKKKinase phosphatases and/or MAPK phosphatases are down-regulated and
inhibited for MAPK activation, because MAPKKKinase introduced experimentally fails to
properly activate MAPK on its own \cite{verlhac00}. Similar results have been obtained 
in {\it Xenopus} oocytes \cite{beaujois09}. As proposed in in mice oocytes Mos may intervene 
at one of the downstream levels by 
affecting the phosphorylation/dephosphorylation equilibrium in a concentration-dependent
manner, by favouring phosphorylation such that the feedforward loop stabilizes $z_3$,
as sketched in Figure 2 B).

In the context of our phenomenological model this shift of the phosphorylation equilibria 
leads to an effective concentration dependence of the kinetic parameters $V_9$ and $V_{10}$. 
We propose therefore the replacement of the reaction constants
\begin{equation}
V_9 \rightarrow V_9(x) = \frac{V_9^*}{K^* + x}\,,\,\,\, 
V_{10} \rightarrow V_{10}(x) = \frac{V_{10}^*}{K^* + x}\,.
\end{equation}
This modification of the reaction constants increases the cooperativity 
of the network, as characterized by the Hill coefficient,  from four to six.
The reason for this is the inverse polynomial dependence we assumed
for simplicity. The same effect would be brought about by a similar intervention 
at the level of MEK.

This result has an important consequence on the idea of  `modularity' of the cascade.
As argued in the introduction, the ubiquitous presence of the MAPK signaling pathway and
its nonlinear characteristics make it attractive to consider it as a recurrent `invariable module' that
is programmed at cascade entry, and whose output interacts with other network elements.
Our simple argument above reveals that an intervention {\it inside} the cascade, by
affecting the phosphorylation/dephosphorylation equilibrium, can have a significant
effect on the output (a more nonlinear signal). In this context, it would not make sense anymore
to consider the MAPK cascade as an invariable modular element. 

\section{Bistability: reconsidering the de Angeli-model}

The foregoing discussion has established how a gradual signal of Mos, $x$, 
is transformed into a response downstream of the cascade. We now discuss what happens 
when this nonlinear signal transformation is embedded into a circuit via a feedback loop, 
see Figure 2 C). In this loop, the cascade output interacts back on the input, $x$,
and therefore favors the production of $x$, autoactivating the cascade.

In order to implement the feedback loop, one has to postulate a dynamics
of Mos. Angeli et al. chose the phenomenological expression \cite{angeli04}.
\begin{equation} \label{mos1}
\dot{x} = -\gamma\frac{x}{K_2 + x} + \tilde{V}_0 z_3(x) x + \tilde{V}_1\,,
\end{equation}
In this model, the first term describes the degradation of Mos with a Michaelis-Menten
kinetics. The second term gives the feedback activation of Mos by MAPK, the output
of the cascade. The function $z_3(x)$ can be calculated exactly for both the linear
kinetics as well as for the Michaelis-Menten kinetics, as shown before. Finally, the
last term in the equation describes a basal production of Mos due to translation from 
its mRNA. We note that the parameters in this equation have been renamed with respect 
to the equations before, taking into account that we
had normalized the variables previously. 

In a subsequent erratum published on their website, 
the authors modified this equation by changing the feedback term into 
$ \tilde{V}_0 z_3(x) $. We now discuss the consequences of this change. 
We start out from a yet slightly modified version of the Mos-dynamics, namely
\begin{equation} \label{mos2}
\dot{x} = -\gamma x + \tilde{V}_0 z_3(x) x\,,
\end{equation}
i.e., we include only a linear degradation term in $x$, and drop the 
constant production
term $\tilde{V}_1$, since it has no qualitative effect on the dynamics.

Figure 5 (top) shows the two curves $\gamma x $ and $\tilde{V}_0 z_3(x) x $ 
whose intersections determine the equilibria. 
It can be seen that there are at most two
intersections, not three. This is due to the fact that for large $x$, 
since $z_3(x)$ saturates, both curves behave as linear functions with 
$z_3$ turning linear {\it from below}. Thus, there cannot be three solutions.
This is even more apparent if one determines the mechanical potential $W$ 
which is easy to do now, since the dynamics in one-dimensional, i.e., we have
\begin{equation}
\dot{x} = - \frac{dW}{dx}\, .
\end{equation}
The potential $W(x)$ is shown in Figure 5 (bottom). It shows either a single 
stable minimum at $ x = 0 $ or a stable minimum at $ x = 0 $ and an unstable 
maximum at a value $ x = x_{um} $. 

The effect of the modification of the feedback term made 
by Angeli {\it et al.} now becomes clear: it is needed in order to obtain three 
intersections, and hence an additional, stable minimum in $W(x)$. However, 
the mathematical expression given by Angeli {\it et al.} still has problems. First, it is not
in accord with what is shown in Fig 5 c of their paper, since there a 
linear curve and the curve $z_3(x)$ are superimposed. This would correspond 
to our equation (\ref{mos2}) with the production term modified as suggested 
by Angeli {\it et al.} However, this modification leads to yet another problem: it means 
that doubly-phosphorylated MAPK can act as a source for Mos, since $\dot{x} \sim z_3 $.

In order to be in  qualitative accord with the results deduced by Angeli {\it et al.} 
one can therefore suggest a phenomenological expression
\begin{equation} \label{mos3}
\dot{x} = -\gamma x + \frac{\tilde{V}_0 z_3(x) x}{\tilde{K} + x}\,
\end{equation}  
which indeed does the give the three intersections, and hence two stable and
an unstable solutions, as can be seen in Figure 6, top and bottom.

The origin of this modeling ambiguity is easy to understand. If there is only one
type of Mos considered, it has to be an active form, capable of initiating the cascade. 
On the other hand, if it is MAPK, $z_3$, which activates Mos, there is evidently
one species lacking, non-activated Mos. For this, one has to properly distinguish 
between two active forms, only one of which is stable. In order to make the Angeli {\it et al.} 
model more biologically realistic, therefore at least a
distinction between a stable and a non-stable form of Mos is needed.  Within the
Angeli model, this can be implemented on the level of the degradation of Mos.
We therefore close the paper by showing how the distinction between stable and non-stable
Mos can lead to a switching of the cascade by acting on the degradation rate, $\gamma$. 

The biological origin of this mechanism is easily justified, see the illustration in Figure 7. 
At early times, non-stable Mos is produced from its mRNA stock and degraded by the 
proteasome. Upon phosphorylation, Mos is shielded against degradation. Indeed, 
Mos stability is brought about by phosphorylation which leads to changes in the degradation 
rate of Mos \cite{sheng02}. Two types of kinases have been shown to be responsible for 
phosphorylation site modification on residues essential to Mos protein stability. These
proteins are either components of the MAPK network or components of the M-phase
Promoting Complex (MPF) \cite{matten96,castro01}. Changes in Mos phosphorylation
have been observed at two key steps in the oocyte cell cycle: (1), during meiosis
resumption (G2/M cell cycle transition, here denoted by $t_1$) and, (2), upon fertilization
(the metaphase-anaphase transition, here denoted by $t_2$). During the time
interval $t_2 - t_1$, Mos is shielded against degradation, allowing its accumulation 
from its mRNA stock (note that this interval is on a timescale of 24 hours). Upon 
dephosphorylation the system goes back to a high degradation rate, driving the 
disappearance of Mos and the inactivation of Mos activity within 30 minutes
\cite{bodart02,castro01}.

When fertilization occurs, Mos is dephosphorylated and
the system goes back to an effectively high degradation rate. Note that the interval $t_2 - t_1$ is on
the timescale of 24 h, while the transition from high to low degradation is much faster (30 min).

This mechanism is easily implemented in the phenomenological model. As shown in Figure 6, 
if the degradation rate is high, the two curves $\gamma x $ and $ \sim z_3(x)x/(\tilde{K} + x) $ 
do not have any common intersection other than at zero: the switch is off. Upon a decrease 
in the degradation rate both curves can intersect, giving rise to  the three intersections
discussed before. Coming back to our model, eq.(\ref{mos3}), we recognize that
it is indeed simplified in the sense that we had omitted a source for Mos, $x$,
which is independent from MAPK. We now put it back in and neglect the presence of the 
MAPK-induced Mos production because all we need in order to realize the above scenario is
a mechanism to modulate Mos concentration from low values (which corresponds to
an effective high degradation rate) to high values (low rate of degradation) back to
low values. 

The ODE with an effective time-dependent degradation rate which 
contains a time-varying production and a constant degradation term 
has the very simple form
\begin{equation}
\dot{x} = - \gamma_{eff}x \equiv (m(t) - \gamma) x\, .
\end{equation}
where we suppose that for all times, $ m(t) \leq \gamma $.
The experimental observations allow to conclude a variation of $m(t)$ such that
\begin{equation}
\gamma_{eff}(t) = \gamma_1\,,\,\,\, t < t_1\,,\,\,\, t > t_2 
\end{equation}
and
\begin{equation}
\gamma_{eff}(t) =  \gamma_2\,,\,\,\, t_1 < t < t_2 
\end{equation}
with $\gamma_1 > \gamma_2 $.  We thus obtain a switching-on and -off of the cascade.

Since the change in the effective degradation rate is fast, the slope of the degradation
curve changes quickly from a steep to a flat profile. Therefore, the barrier
separating the two stable states becomes shallow and its location moves close
to the minimum of $ W(x) $ at low concentration of Mos (see Figure 6). This
switch enables the system to quickly exit from a now unstable state.

\section{Conclusion}

In this paper we have presented a discussion of a phenomenological model
of the MAPK cascade. We have shown that the equilibrium states of the MAPK cascade 
can be determined exactly even for a model with Michaelis-Menten kinetics,  
a fact to our knowledge so far overlooked.

Linear and Michaelis-Menten kinetics lead to identical asymptotic Hil coefficients.
The comparison of Hill coefficients from theory to experimental data, however, 
must be taken with caution, since the values seen in experiment are essentially effective 
exponents, i.e. most likely not in the proper asymptotic regime.  In this respect it is
therefore very important whether a linear or Michaelis-Menten kinetics is assumed.

Further, we have shown how a simple feedforward mechanism acting on the inner levels 
of the cascade can increase the cooperativity of the cascade. This is an important result 
since it means that a `simple' rewiring of the cascade can increase cooperativity without the
addition of new cascade levels or phosphorylation steps.

This result has two important consequences. The first is obviously again for experiment, since
a high measured value of the Hill coefficient might need an explanation in terms of a more
detailed study of the interactions in the network; such study is currently under way
\cite{beaujois09}. The second consequence arises for the notion of modularity. In its strongest
interpretation one might wish to consider the MAPK cascade as an {\it invariable} module 
to be implemented in a network all `around it'. In order to optimize the response, the 
feedforward mechanism we propose can be advantageous, since it increases the 
cooperativity of the cascade, but it does so at an expense of cascade modularity.
Our finding supports the idea that it is best to understand the cascade as a particular 
motif within a network: a motif which has no boundaries at any level to molecular 
interactions across the whole network.

Finally, we pointed out that the phenomenological modeling of the feedback loop
which turns the MAPK cascade into a switch by selecting two out of an infinitude of possible 
states of the cascade needs to be done with caution. We have put forward a discussion which 
addresses some problems present in the model and its solution as presented by Angeli {\it et al.} 
The origin of this ambiguity is the fact that the model is too reduced: a proper distinction
between the different states of Mos is needed. We have argued that the Angeli model needs
at least a time-dependent degradation constant in order to properly account for the
cascade dynamics.
\\

{\bf Acknowledgment.} CR gratefully acknowledges support through a 
grant from the French National Cancer Institute, INCa.
\\

\newpage
 
\begin{figure}[h]
\includegraphics[width=8cm,height=6.4cm,angle=0]{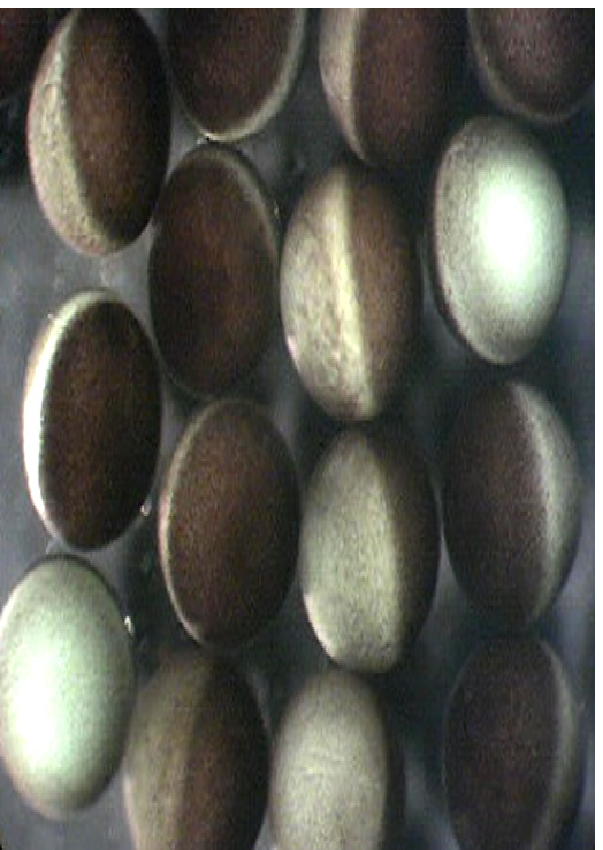}
\caption{{\it Xenopus} oocytes. The cells are large: a typical diameter is
about 1 mm.}
\end{figure}

\begin{figure}
\includegraphics[height=14cm,angle=0]{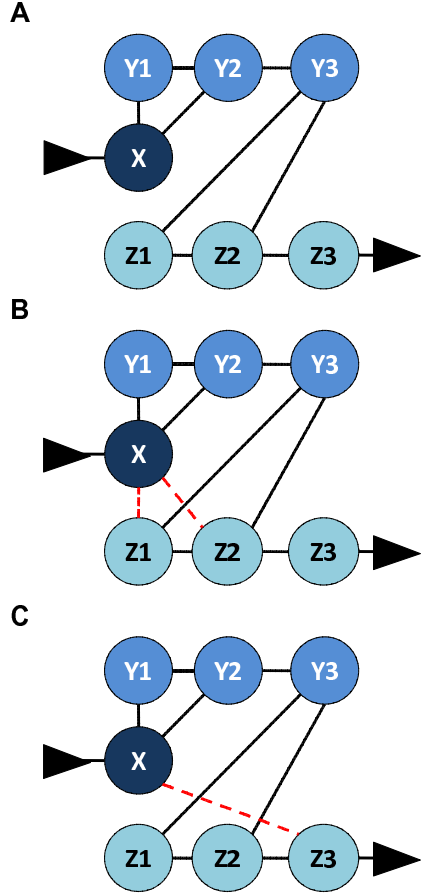}
\caption{(Color online.) Schematics of the MAPK cascade in the simplified model
proposed in ref.\cite{angeli04} and extended here. The nomenclature is as follows:
$x$ Mos, $y_{1,2,3}$ MEK of different, increasing phosphorylation status,
and likewise $z_{1,2,3}$ for MAPK. A) shows the basic cascade: Mos promotes
the phosphorylation of MEK, doubly-phosphorylated MEK, $y_3$, promotes the 
phosphorylation of MAPK. B) Mos promotes the phosphorylation of MAPK in 
an indirect way, as discussed in the text; C) the cascade with feedback of $z_3$ on $x$.}
\end{figure}

\begin{figure}
\includegraphics[height=9cm,angle=0]{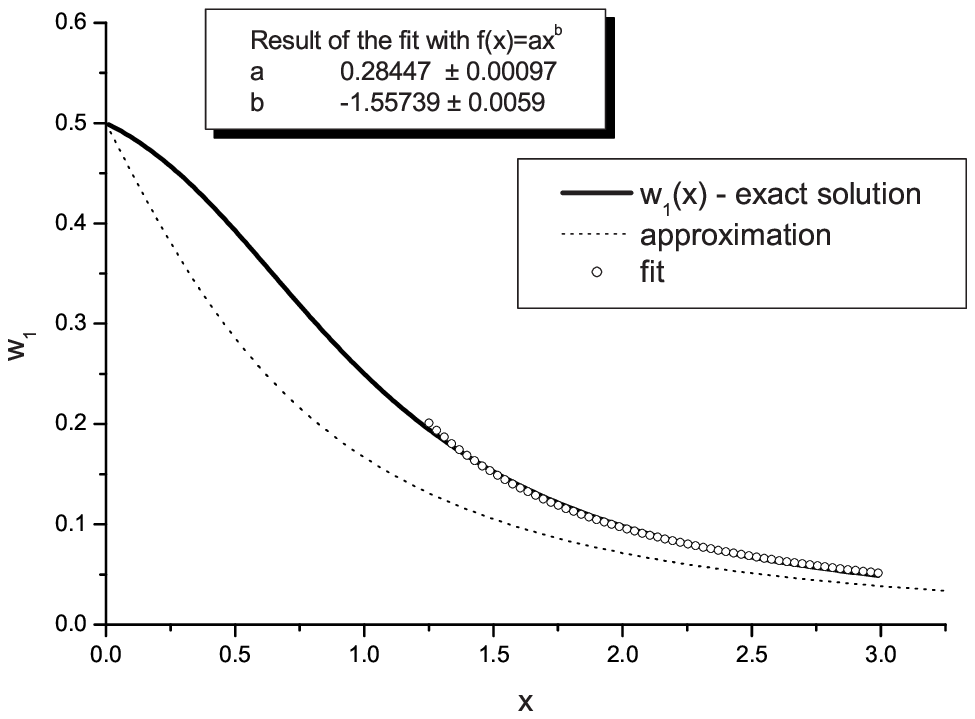}
\includegraphics[height=9cm,angle=0]{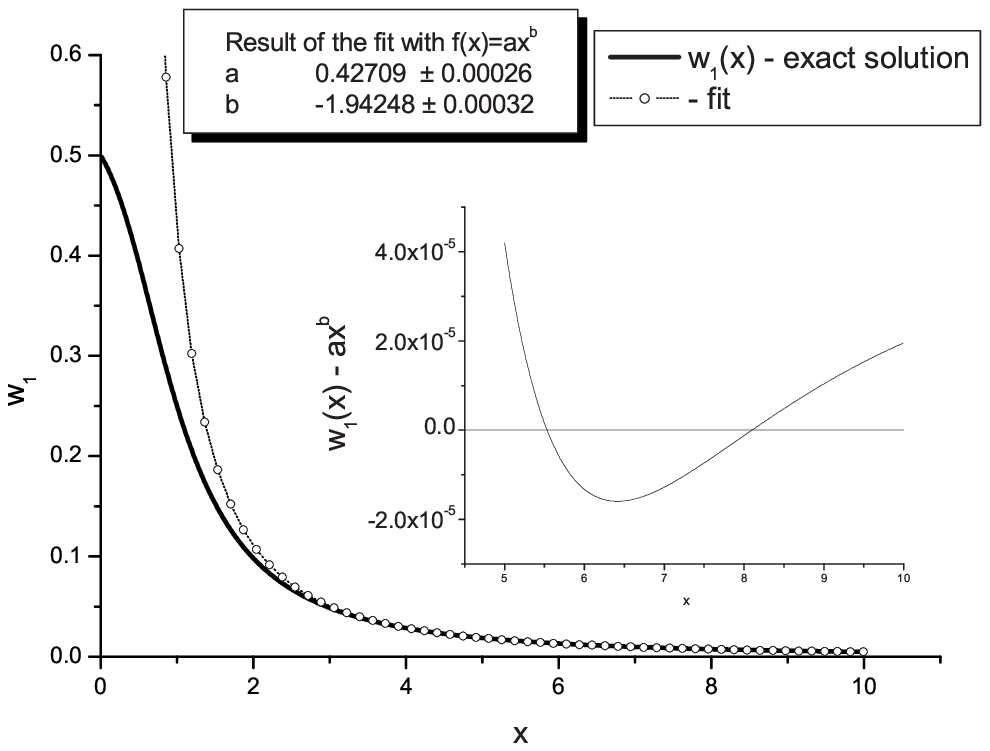}
\caption{The concentration of $w_1$ (the transformed concentration of 
non-phosphorylated MEK, $y_1$) as a function
of concentration of Mos, $x$. The Hill exponent deviates from a value of
two over the accessible concentration ranges; the asymptotic regime, for
which this value would be reached, is out of range, see the discussion
in the main text.}
\end{figure}

\begin{figure}
\includegraphics[height=8cm,angle=0]{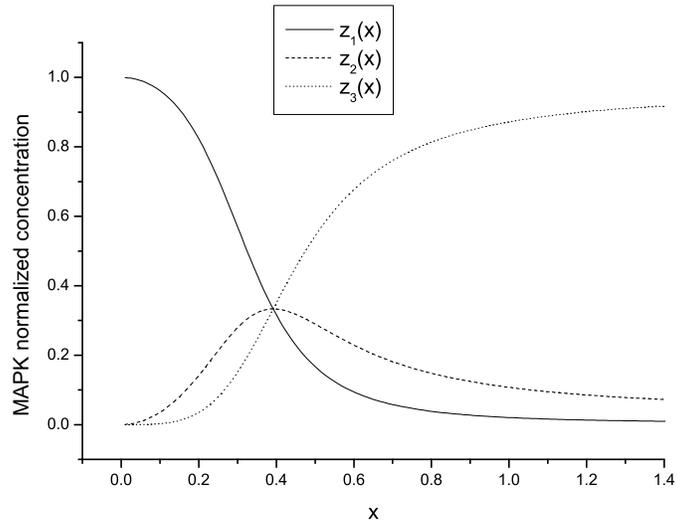}
\caption{The dependence of $z_1$ and its phosphorylated forms $z_2$ and $z_3$
as obtained from the full solution to equation (\ref{conservation_w1}), using
the stationary state conditions, and transforming back to the original
variables.}
\end{figure}

\begin{figure}
\includegraphics[height=8cm,angle=0]{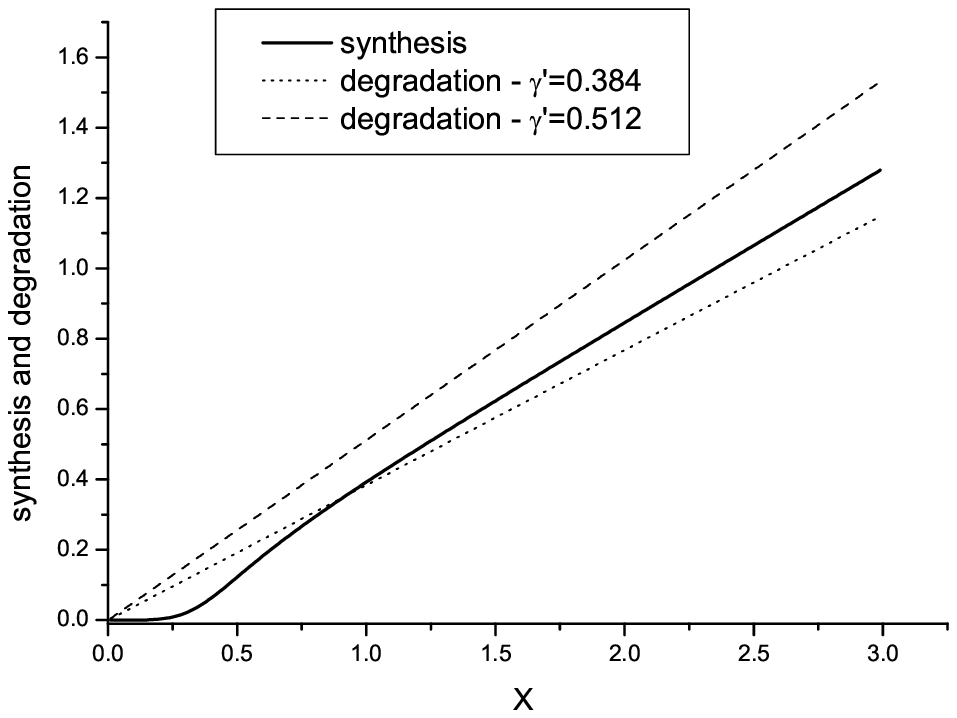}
\includegraphics[height=8cm,angle=0]{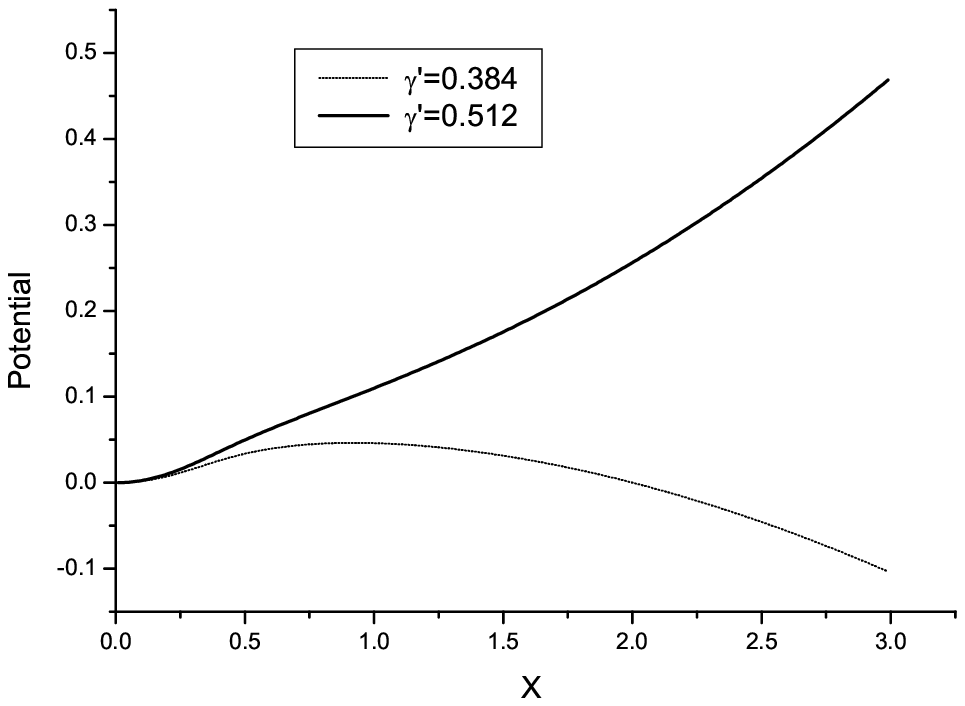}
\caption{Top: stationary state conditions for the model by Angeli et al.
Two intersections correspond to a stable and an unstable state. See
Bottom: potential.}
\end{figure}

\begin{figure}
\includegraphics[height=8cm,angle=0]{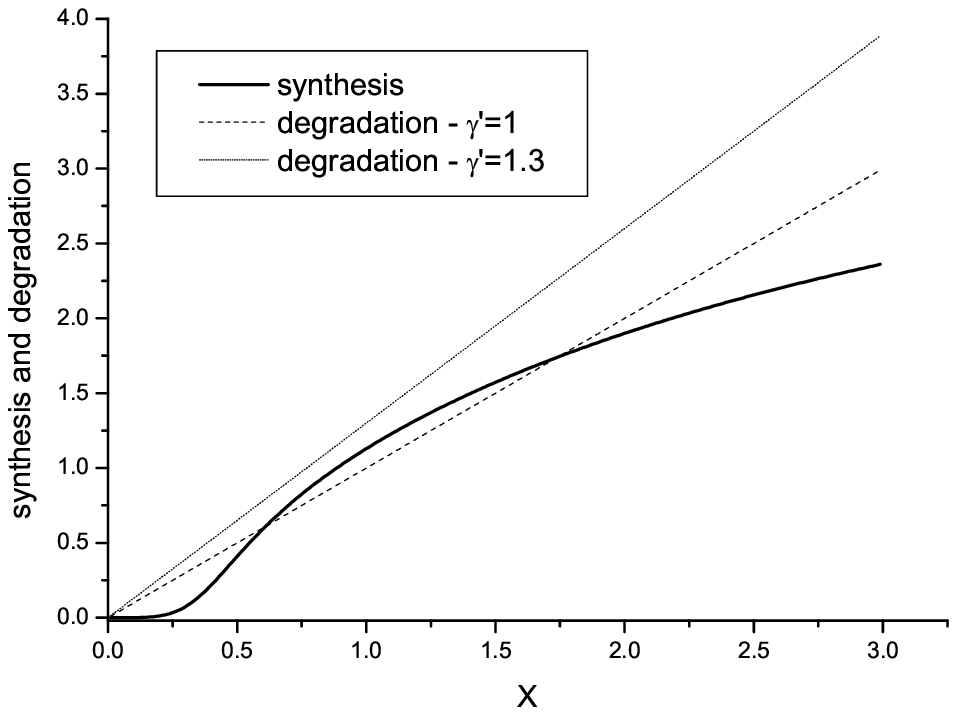}
\includegraphics[height=8cm,angle=0]{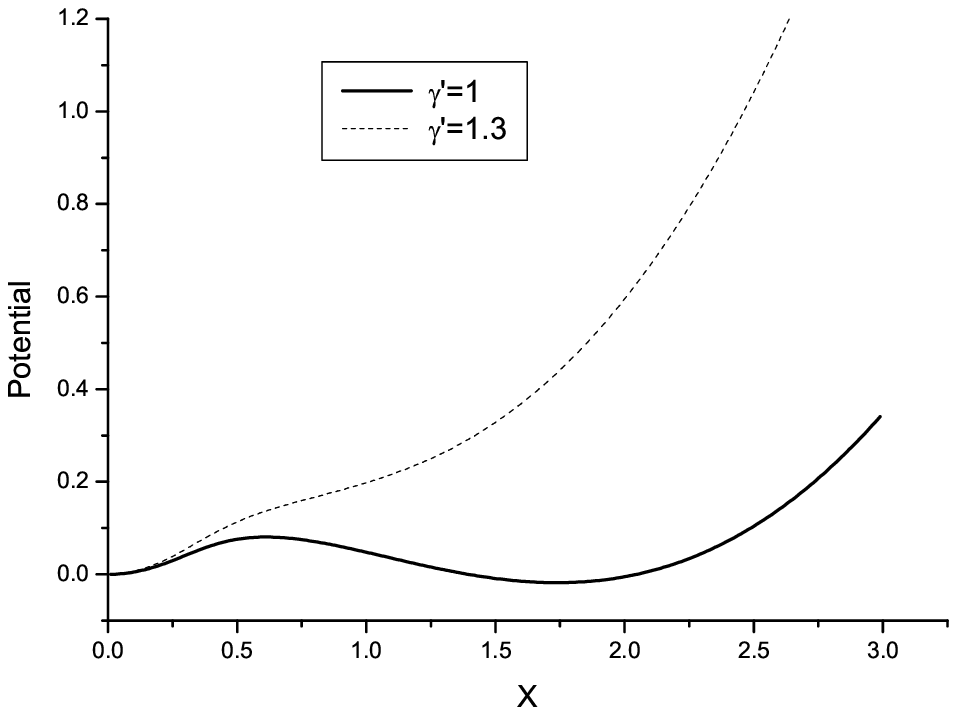}
\caption{Top: stationary state conditions for the model given by eq.(\ref{mos3}). 
Three intersections exist. See Bottom: potential.}
\end{figure}

\begin{figure}
\includegraphics[height=6cm,angle=0]{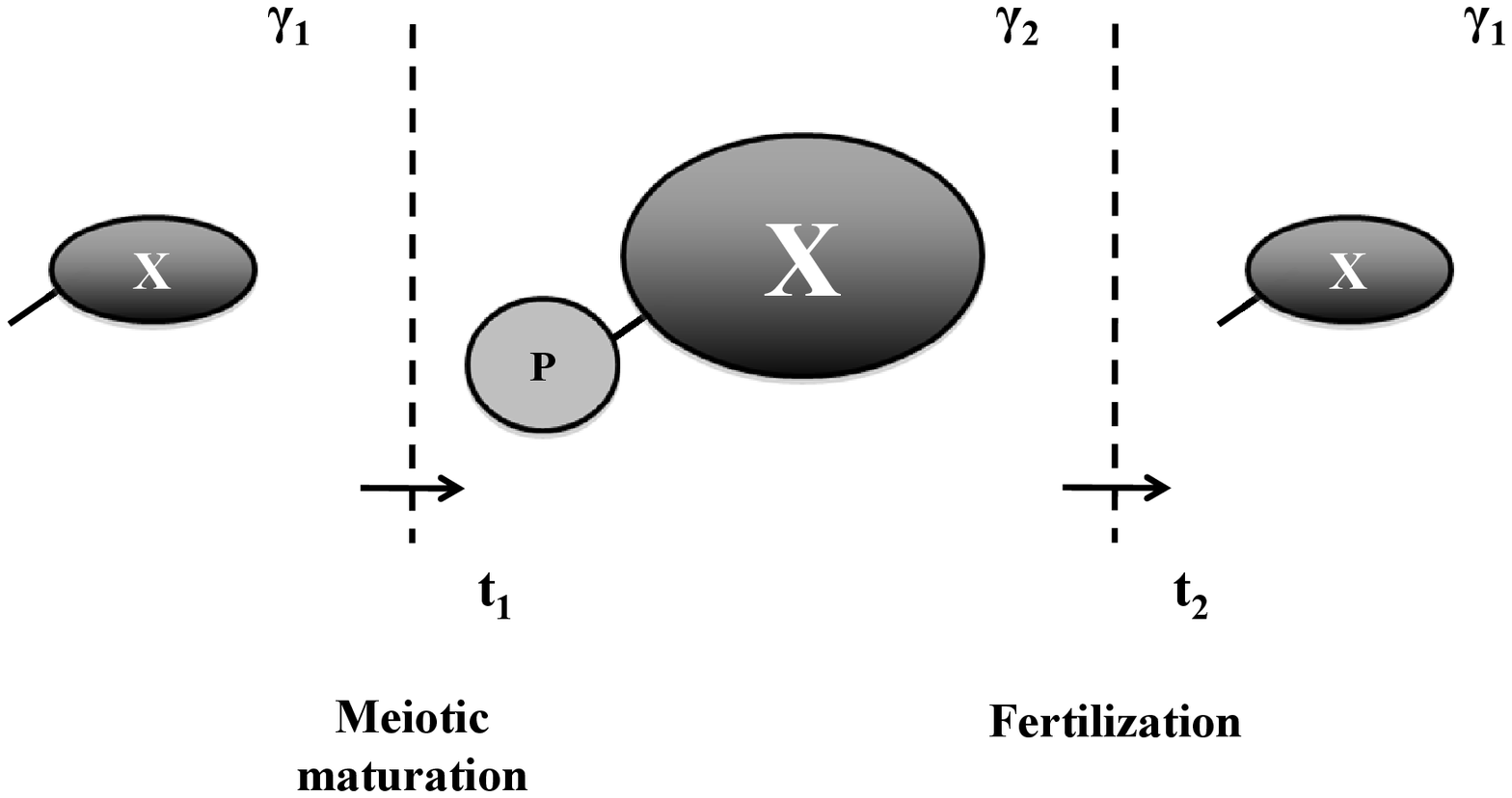}
\caption{The switching mechanism by modulating Mos-degradation.}
\end{figure}


\newpage

\begin{thebibliography}{99}
\bibitem{gomperts02} B.~D.~Gomperts, I.~M.~Kramer and P.~E.~R. Tatham 
{\it Signal Transduction}, Academic Press, San Diego (2002)
\bibitem{ferrell96} J.E.~FerrellNJr., Trends Biochem. Sci. {\bf 21}, 460 (1996)
\bibitem{ferrell01} J.E.~Ferrell~Jr. and W.~Xiong, Chaos, {\bf 11}, 227 (2001)
\bibitem{tyson03} J.~J.~Tyson, K.~C.~Chen and B."ovak, Curr. Op. Cell Biol. 
{\bf 15}, 221
\bibitem{sagata88} N. Sagata {\it et al}, Nature {\bf 335}, 519 (1988) 
\bibitem{huang96} C.-Y.~F.~Huang and J.~E.~Ferrell,~Jr., PNAS {\bf 93},
10078 (1996)
\bibitem{bluethgen01} N.~Bl\"uthgen and H.~Herzel, in
{\it 2nd Workshop on Computation of Biochemical Pathways and
Genetic Networks} - Berlin: Logos, 55 (2001)
\bibitem{markevich04} N.~I.~Markevich, J.~B.~Hoek and B.~N.~Kholodenko,
Journ. Cell Biology {\bf 164}, 353 (2004)
\bibitem{gunawardena05} J.~Gunawardena, PNAS {\bf 102}, 14617 (2005)
\bibitem{xiong03} W.~Xiong, J.~E.~Ferrell, Jr., Nature {\bf 426}, 460 
(2003)
\bibitem{kholodenko00} B.~N.~Kholodenko, Eur. J. Biochem. {\bf 267},
1583 (2000)
\bibitem{wang06} X.~Wang, N.~Hao, H.~G.~Dohlmann and T.~C.~Elston,
Biophys. J. {\bf 90}, 1961 (2006)
\bibitem{markevich06} N.~I.~Markevich, M.~I.~Tayganov, J.~B.~Hoek
and B.~N.~Kholodenko, Mol. Sys. Biol. doi:10.1038/msb4100108 (2006)
\bibitem{ventura08} A.~C.~Ventura, J.-A.~Sepulchre and S.~D.~Merajver, PLoS Comp.
Biology {\bf 4}, e1000041 (2008) 
\bibitem{angeli04} D.~Angeli, J.~E.~Ferrell,~Jr., and E.~D.~Sontag,
PNAS {\bf 101}, 1822 (2004)
\bibitem{alon06} U.~Alon, {\it An Introduction to Systems Biology},
CRC Press (2006)
\bibitem{verlhac00} M.~H.~Verlhac et al., EMBO J. {\bf 19}, 6065 (2000)
\bibitem{beaujois09} R.~Beaujois and J.F. Bodart, unpublished (2009)
\bibitem{bodart02} J.-F.~Bodart, S.~Flament and J.-P.~Vilain, {\it Mol. Reprod. Dev.}
{\bf 61}, 570 (2002)
\bibitem{sheng02}  J.~Sheng, A.~Kumagai, W.G.~Dunphy and A.~Varshavsky, {\it EMBO J.}Ê21, 6061 (2002)
\bibitem{castro01} A.~Castro et al., {\it Mol. Biol. Cell.} {\bf 12}, 2660 (2001)
\bibitem{matten96} W.T.~Matten, T.D.~Copeland, N.G. Ahn and G.F.~Vande Woude, 
{\it Dev. Biol.} {\bf 179}, 485 (1996)
\end{thebibliography}
\end{document}